\documentclass[cameraready]{Interspeech}

\title{Absorbing Discrete Diffusion for Speech Enhancement}

\author[orcid=0009-0006-4965-3514]{Philippe}{Gonzalez}
\address{Department of Health Technology, Technical University of Denmark}
\email{phigon@dtu.dk}


\usepackage{lipsum}
\usepackage{subcaption}
\usepackage{microtype}
\usepackage{cite}

\DeclareMathOperator*{\argmin}{arg\,min}

\begin{document}

\maketitle

\begin{abstract}
Inspired by recent developments in neural speech coding and diffusion-based language modeling, we tackle speech enhancement by modeling the conditional distribution of clean speech codes given noisy speech codes using absorbing discrete diffusion.
The proposed approach, which we call \textit{ADDSE}, leverages both the expressive latent space of neural audio codecs and the non-autoregressive sampling procedure of diffusion models.
To efficiently model the hierarchical structure of residual vector quantization codes, we propose \textit{RQDiT}, which combines techniques from RQ-Transformer and diffusion Transformers for non-autoregressive modeling.
Results show competitive performance in terms of non-intrusive objective metrics on two datasets, especially at low signal-to-noise ratios and with few sampling steps.
Code and audio examples are available online\footnote{\url{https://philgzl.com/addse-demo}}.
\end{abstract}

\section{Introduction}

Diffusion models have recently gained attention as powerful generative approaches for speech enhancement (SE)~\cite{lu2022conditional,richter2023speech,gonzalez2024investigating,lemercier2025diffusion}.
They typically consider a conditional probability distribution over clean speech given noisy speech in the continuous short-time Fourier transform (STFT) domain.
However, the high dimensionality of the STFT representation, coupled with the iterative nature of the sampling procedure, leads to high computational demands.
In computer vision, diffusion models often operate in the latent space of a variational autoencoder~\cite{kingma2014auto,rombach2022high,peebles2023scalable,chen2024pixart,karras2024analyzing}.
This transposes the generative modeling task to a well-behaved and perceptually relevant space with fewer dimensions.
Meanwhile, only a handful of studies have explored latent diffusion for SE~\cite{dhyani2025high,kumar2025prose,guimaraes2026ditse}.

Concurrently, neural audio codecs (NACs) have been widely adopted for low-bitrate speech coding~\cite{zeghidour2021soundstream,defossez2023high,kumar2023high,xin2024bigcodec,defossez2024moshi,della2025focalcodec}.
They have also been applied to downstream tasks successfully, including SE~\cite{wang2024selm,xue2024low,yang2024genhancer,li2024masksr,liu2025joint,fu2026discontse,li2025speech,yao2025gense,kang2025llase,liu2025universal,kammoun2026modeling,lanzendorfer2026high,della2025autoregressive,yan2025unise}.
Notably, the discrete nature of the NAC representation allows applying Transformer-based language modeling techniques seamlessly.
However, most approaches rely on next-code prediction and thus autoregression, which leads to slow inference.
Recently, discrete diffusion models~\cite{austin2021structured,campbell2022continuous,meng2022concrete,sun2023score,lou2024discrete,shi2024simplified,ou2025your} have been proposed for non-autoregressive sequence modeling in discrete spaces.
This enables efficient parallel generation while still leveraging the powerful modeling capabilities of Transformers.

The present study tackles SE by modeling the conditional distribution of clean speech codes given noisy speech codes using absorbing discrete diffusion (ADD)~\cite{lou2024discrete,shi2024simplified,ou2025your}.
The proposed approach, which we call \textit{ADDSE}, leverages both the expressive NAC latent space and the non-autoregressive diffusion sampling.
To the best of our knowledge, this is the first study to apply ADD for SE.
While some studies have applied MaskGIT~\cite{chang2022maskgit}, which resembles ADD, for SE~\cite{yang2024genhancer,li2024masksr,liu2025joint,fu2026discontse}, MaskGIT lacks a theoretically grounded formulation, and thus may not approximate a principled likelihood~\cite{shi2024simplified}.
Moreover, these studies did not investigate the time-independent modeling for efficient sampling~\cite{ou2025your}.

\begin{figure}[t]
  \centering
  \includegraphics[scale=0.958]{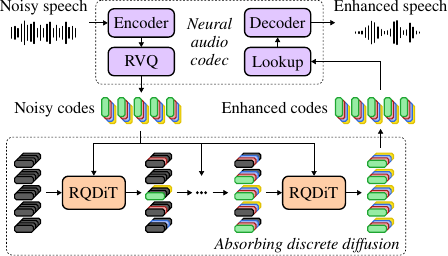}
  \caption{Proposed ADDSE framework.}
  \label{fig:addse}
\end{figure}

\section{Background}

\subsection{Neural audio codecs and residual vector quantization}

A typical NAC~\cite{zeghidour2021soundstream,defossez2023high,kumar2023high} consists of an encoder, a quantizer, and a decoder.
The encoder transforms the input waveform into a low-frame-rate latent representation $\bm{h} \in \mathbb{R}^{L \times H}$, where $L$ is the number of time frames, and $H$ is the dimension of the latent space.
The quantizer then maps the latent representation to discrete codes.
The current dominant approach is residual vector quantization (RVQ)~\cite{zeghidour2021soundstream}, which involves cascading $D$ quantizers.
Each quantizer consists of a learnable codebook with $K$ entries, and takes the residual of the previous quantizer as input.
Formally, the discrete code $c^{i,j} \in \{1, \dots, K\}$ at frame $i \in \{1, \dots, L\}$ and depth $j \in \{1, \dots, D\}$ is given by
\begin{gather}
  c^{i,j} = \argmin_{k \in \{1, \dots, K\}} \lVert \bm{r}^{i,j-1} - \bm{e}(k;j) \rVert_2^2, \\
  \bm{r}^{i,j} = \bm{r}^{i,j-1} - \bm{e}(c^{i,j};j),
\end{gather}
where $\bm{e}(k;j) \in \mathbb{R}^H$ is the $k$-th entry of the $j$-th codebook, $\bm{r}^{i,j} \in \mathbb{R}^H$ is the residual at frame $i$ and depth $j$, and $\bm{r}^{i,0} = \bm{h}^i \in \mathbb{R}^H$ is the latent at frame $i$.
The RVQ output is the discrete code sequence $\bm{c} \in \{1, \dots, K\}^{L \times D}$.
Codes can then be mapped back to the latent space by codebook lookup and summation:
\begin{equation}
  \hat{\bm{h}}^i = \sum_{j=1}^D \bm{e}(c^{i,j};j).
\end{equation}
The decoder then transforms the quantized latent representation $\hat{\bm{h}} = (\hat{\bm{h}}^1, \dots, \hat{\bm{h}}^L) \in \mathbb{R}^{L \times H}$ into the reconstructed waveform.

\subsection{Absorbing discrete diffusion}
\label{sec:add}

We consider the finite support $\{1, \dots, K + 1\}$ over which probability distributions are described by vectors in $[0, 1]^{K+1}$ whose elements sum to 1.
A continuous-time discrete diffusion process can be defined by evolving an initial distribution $\bm{p}_0 = \bm{p}_{\text{data}}$ from $t=0$ to $t=T$ according to a continuous-time Markov chain given by the following Kolmogorov forward equation~\cite{anderson2012continuous,campbell2022continuous},
\begin{equation}
  \frac{d \bm{p}_t}{d t} = \bm{Q}_t \bm{p}_t,
\end{equation}
where $\bm{Q}_t \in \mathbb{R}^{(K+1) \times (K+1)}$ is the transition rate matrix, which has non-negative off-diagonal entries and zero column sums, such that $\bm{p}_t$ remains a valid probability distribution for all $t$.
In practice, $\bm{Q}_t$ is often chosen to diffuse the probability mass towards an absorbing state or mask $M \in \{1, \dots, K + 1\}$.
Namely,
\begin{equation}
  \bm{Q}_t = \sigma(t) \bm{Q}^\text{absorb} = \sigma(t) \resizebox{0.91\width}{!}{$\begin{bmatrix}
    -1 & 0 & \cdots & 0 & 0 \\
    0 & -1 & \cdots & 0 & 0 \\
    \vdots & \vdots & \ddots & \vdots & \vdots \\
    0 & 0 & \cdots & -1 & 0 \\
    1 & 1 & \cdots & 1 & 0
  \end{bmatrix}$},
\end{equation}
where $\sigma(t) \geq 0$ is the noise schedule and $M$ was set to $K+1$ without loss of generality.
The reverse time process is given by the following reverse transition rate matrix~\cite{kelly1979reversibility,sun2023score},
\begin{equation}
  \bar{Q}_{t}^{m,n} = \begin{cases}
    \frac{p_t^m}{p_t^n} Q_{t}^{n, m}, & n \neq m, \\
    -\sum_{k \neq m} \bar{Q}_{t}^{k, m}, & n = m,
  \end{cases}
\end{equation}
where superscripts denote indexing of matrices and vectors.
Simulating the reversal thus requires estimating the concrete score $\frac{p_t^m}{p_t^n}$~\cite{meng2022concrete}, which can be done using a neural network trained with denoising score entropy~\cite{lou2024discrete}.
Alternatively, the score network can be reparametrized to predict a time-independent conditional distribution over clean data~\cite{ou2025your}.
The following denoising cross-entropy (DCE) can thus be used as the training objective,
\begin{equation}
  \mathbb{E}_{\substack{c \sim \bm{p}_\text{data} \\ \lambda \sim \mathcal{U}_{[0, 1]} \\ c_\lambda \sim \bm{p}_\lambda}} - \frac{1}{\lambda} \delta_{c_\lambda M} \log q_\theta(c \mid c_\lambda),
\end{equation}
where the variable $t$ was changed to $\lambda = 1 - e^{-\int_0^t \sigma(s) ds}$, which is the probability of a code being absorbed in $[0, t]$ during the forward process, $q_\theta$ is the conditional distribution predicted by the neural network, which is often parametrized such that $q_\theta(M \mid \cdot) = 0$, and $\delta$ is the Kronecker delta.

For the multi-dimensional case, each dimension is often treated as an independent diffusion process with the same transition rate matrix, allowing joint distributions to factorize across dimensions.
The DCE reduces to
\begin{equation}
  \mathbb{E}_{\substack{\bm{c} \sim \bm{p}_\text{data} \\ \lambda \sim \mathcal{U}_{[0, 1]} \\ \bm{c}_\lambda \sim \bm{p}_\lambda}} \sum_{c_{\lambda}^i = M} - \frac{1}{\lambda} \log q_\theta(c^i \mid \bm{c}_\lambda).
\end{equation}
For sampling, the Euler or Tweedie $\tau$-leaping methods can be used starting from the fully absorbed state $\bm{c}_T = (M, \dots, M)$.
Under a log-linear noise schedule $\sigma(t) = \frac{T}{T - t}$, both methods lead to the same reverse transition probability~\cite{ou2025your},
\begin{equation}
  \label{eq:sampling}
  p_{s \mid t}^i(c_s^i \mid \bm{c}_t) = \begin{cases}
    \frac{t - s}{t} p_0(c_s^i \mid \bm{c}_t), & c_s^i \neq M, c_t^i = M,\\
    1 - \frac{t - s}{t}, & c_s^i = M, c_t^i = M,\\
    \delta_{c_s^i c_t^i}, & c_t^i \neq M,
  \end{cases}
\end{equation}
where $0 \leq s < t \leq T$ and $p_0$ is approximated with $q_\theta$.

The reparameterization of the score network~\cite{ou2025your} not only facilitates learning by eliminating the need for modeling the time-dependency, but also enables more efficient sampling.
Namely, if no code was unabsorbed during the previous step, then the conditional distribution $q_\theta$ can be reused for the current step.

\section{Absorbing discrete diffusion for speech enhancement}

\subsection{Noisy-speech-conditioned absorbing discrete diffusion}

We propose to perform SE by modeling the conditional distribution of clean speech codes given noisy speech codes using ADD.
The framework, which we call \textit{ADDSE}, is illustrated in Fig.~\ref{fig:addse}.
During training, a frozen NAC transforms the clean and noisy waveforms into discrete codes $\bm{c}$ and $\tilde{\bm{c}}$ in $\{1, \dots, K\}^{L \times D}$, respectively.
Each time frame and codebook depth pair is treated as a separate dimension, leading to $L \times D$ independent diffusion processes.
A neural network is trained to predict conditional probabilities $q_\theta$ over clean codes $\bm{c}$ given noisy codes $\tilde{\bm{c}}$ and partially absorbed clean codes $\bm{c}_\lambda$ with the following DCE,
\begin{equation}
  \label{eq:dce}
  \mathbb{E}_{\substack{\bm{c}, \tilde{\bm{c}} \sim \bm{p}_\text{data} \\ \lambda \sim \mathcal{U}_{[0, 1]} \\ \bm{c}_\lambda \sim \bm{p}_\lambda}} \sum_{c_{\lambda}^{i,j} = M} - \frac{1}{\lambda} \log q_\theta(c^{i,j} \mid \bm{c}_\lambda, \tilde{\bm{c}}),
\end{equation}
where $\bm{p}_\text{data}$ is the training data distribution of noisy and clean codes, and $(i, j)$ are the frame and depth indices of the codes.
During inference, the noisy codes are obtained from the input noisy waveform, and the clean codes are initialized to the fully absorbed state $(M, \dots, M)$ with $M = K + 1$.
Clean codes are then generated by sampling according to~\eqref{eq:sampling}.
Finally, the NAC transforms the generated clean codes back to the time domain.

\subsection{Hierarchical non-autoregressive sequence modeling}

In contrast to traditional sequence modeling where a single sequence is considered, the hierarchical nature of the RVQ codes requires modeling multiple sub-sequences corresponding to different codebook depths.
Flattening the codes and treating them as a single sequence substantially raises computational demands for algorithms with non-linear complexity in sequence length, such as Transformers.
Moreover, it does not leverage the hierarchical structure of the codes.
To address this, RQ-Transformer was proposed~\cite{lee2022autoregressive}, which comprises two Transformers applied along the space and depth dimensions, respectively.

\begin{figure}[t]
  \centering
  \begin{subfigure}{\linewidth}
    \centering
    \includegraphics[scale=0.96]{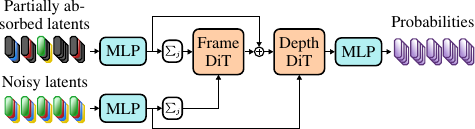}
    \caption{RQDiT}
    \label{fig:rqdit}
  \end{subfigure}\\[6pt]
  \begin{subfigure}{\linewidth}
    \centering
    \includegraphics[scale=0.96]{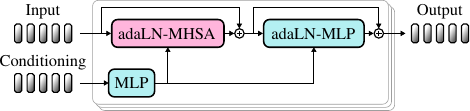}
    \caption{Frame- or depth-DiT}
    \label{fig:dit}
  \end{subfigure}
  \caption{
    RQDiT architecture.
    Two DiTs are applied along the NAC frame and depth dimensions, respectively.
  }
  \label{fig:rqdit_dit}
\end{figure}

Inspired by this, we adapt RQ-Transformer for non-autoregressive sequence modeling, by predicting probabilities over clean codes $q_\theta(c^{i,j} \mid \bm{c}_\lambda, \tilde{\bm{c}})$ for reverse ADD, instead of predicting the code at the next position.
This is similar to diffusion Transformers (DiT) for continuous data~\cite{peebles2023scalable}.
The resulting architecture, which we thus call \textit{RQDiT}, is illustrated in Fig.~\ref{fig:rqdit_dit}.
Instead of learning an embedding layer at the input of RQDiT to map discrete codes to continuous vectors, we reuse the codebook entries from the NAC $\bm{e}(c^{i,j}_\lambda; j)$ and $\bm{e}(\tilde{c}^{i,j}; j)$, which are already optimized for representing the latent space.
Since the mask code $M$ does not correspond to any codebook entry, we assign it the zero vector, i.e.\ $\bm{e}(M; j) = \bm{0}$ for all $j$.
Multilayer perceptrons (MLPs) map the selected entries to the hidden dimension of two DiTs.
The first DiT is applied along the frame dimension and processes the sum over codebook depths similar to RQ-Transformer.
The output of the frame-DiT is added to the hidden input activations at each codebook depth.
The second DiT then processes each time frame independently along the depth dimension.
An MLP finally maps the output of the depth-DiT to probabilities over clean codes $q_\theta(c^{i,j} \mid \bm{c}_\lambda, \tilde{\bm{c}})$ for each frame and depth pair.
Conditioning on the noisy codes $\tilde{\bm{c}}$ is done using adaptive layer normalization (adaLN)~\cite{perez2018film,peebles2023scalable} in both DiTs.
Each DiT comprises multi-head self-attention (MHSA) blocks and MLPs with residual connections in multiple layers.
We use rotary position embedding (RoPE)~\cite{su2024roformer} for encoding the frame and depth positions in the frame- and depth-DiT, respectively.

\section{Experimental setup}

\subsection{Datasets}
\label{sec:datasets}

Noisy speech is simulated dynamically during training~\cite{zeghidour2021wavesplit,choi2022empirical} by mixing clean speech utterances with noise segments at \qty{16}{\kilo\hertz}.
Clean speech utterances are selected from DNS5~\cite{dubey2024icassp}, LibriSpeech~\cite{panayotov2015librispeech}, MLS~\cite{pratap2020mls}, VCTK~\cite{veaux2013voice}, and EARS~\cite{richter2024ears}.
Noise segments are selected from DNS5~\cite{dubey2024icassp}, WHAM!~\cite{wichern2019wham}, FSD50K~\cite{fonseca2022fsd50k}, FMA~\cite{defferrard2017fma}, and DEMAND~\cite{thiemann2013demand}.
The signal-to-noise ratio (SNR) is uniformly selected between \qtylist{-5;15}{\decibel}.
For testing, we build two datasets each comprising \num{1000} mixtures.
\textit{Libri-TUT} combines clean speech utterances from the held-out test split of LibriSpeech with noise segments from TUT~\cite{mesaros2016tut}.
No segments from TUT are used during training, and thus this dataset evaluates cross-corpus noise generalization.
\textit{Clarity-FSD50K} combines clean speech utterances from Clarity~\cite{cox2022clarity} with the held-out test split of FSD50K.
Similarly, no utterances from Clarity are used during training, and thus this dataset evaluates cross-corpus speech generalization.
Moreover, FSD50K comprises a broad range of sound events, while TUT is limited to relatively stationary urban recordings.
Both datasets thus represent significantly different acoustic conditions.

\subsection{Neural audio codec}

We train our own NAC following the popular mirrored convolutional encoder-decoder architecture from previous work~\cite{zeghidour2021soundstream,defossez2023high,kumar2023high,xin2024bigcodec,defossez2024moshi}, with the same training data as described in Sec.~\ref{sec:datasets}.
We use Snake activation functions~\cite{ziyin2020neural} as in~\cite{kumar2023high} and long short-term memory (LSTM) cells in each layer as in~\cite{xin2024bigcodec}.
The downsampling factors in each encoder layer are set to 2, 2, 4, 4, and 5, leading to an overall downsampling factor of 320 and a frame rate of \qty{50}{\hertz}.
The RVQ uses $D=4$ codebooks with $K=1024$ entries each, leading to a bitrate of \qty{2}{kbps}.
In contrast to state-of-the-art speech NACs which achieve very low bitrates by using a single codebook~\cite{della2025focalcodec}, we find that using multiple codebooks is crucial for achieving good reconstruction quality, which we attribute to the wide range of noise types in our training data.
We use the low-dimensional $\ell_2$-normalized codebook lookup trick from~\cite{yu2022vector} as in~\cite{kumar2023high}.
We use a multi-period discriminator (MPD)~\cite{kong2020hifi} and a multi-scale short-time Fourier transform discriminator (MS-STFT)~\cite{defossez2023high} as in~\cite{xin2024bigcodec}.
The training objective combines an $\ell_1$ multi-scale mel-spectrogram loss with weight 15, an adversarial hinge loss with weight 1, an $\ell_1$ discriminator feature matching loss with weight 1, an $\ell_2$ codebook loss with weight 1, and a commitment loss with weight 0.25.
We train for \num{500000} steps using Adam~\cite{kingma2015adam} with a learning rate of \num{1e-4}, a batch size of 32, and \qty{1}{\second}-long audio segments.
The same NAC is used for encoding both the clean and noisy speech for ADDSE.

\subsection{Compared systems}

We train RQDiT with hidden dimensions 96, 192, 384, 768, and 1152.
The resulting systems are denoted as \textit{ADDSE-XS}, \textit{ADDSE-S}, \textit{ADDSE-M}, \textit{ADDSE-L}, and \textit{ADDSE-XL}, respectively.
The number of layers and attention heads in each DiT are both fixed to 12.
We train for \num{1000000} steps using AdamW~\cite{loshchilov2019decoupled} with a learning rate of \num{1e-4}, a batch size of 16, \qty{4}{\second}-long code sequences, and an $\ell_2$ gradient clipping norm of 1.
The number of uniform sampling steps $N_\text{steps}$ is varied between 1 and 1024.

Two popular discriminative baselines Conv-TasNet~\cite{luo2019conv} and BSRNN~\cite{luo2023music} are trained with signal-to-distortion ratio (SDR).
BSRNN ranked first in the URGENT 2025 Challenge~\cite{sun2025scaling} and is thus considered the state of the art.
We also use the generative baseline SGMSE+~\cite{richter2023speech}, which performs continuous diffusion in the STFT domain.
Additionally, we develop \textit{EDM-SE}, which also performs STFT-domain diffusion, but implements advances from EDM~\cite{karras2022elucidating,karras2024analyzing}.
Namely, it uses the preconditioning, the Heun sampler, the log-normal training noise schedule, the polynomial sampling noise schedule, and the ADM architecture~\cite{dhariwal2021diffusion}.
The preconditioning and the sampler were already shown to improve SE performance~\cite{gonzalez2024diffusion,gonzalez2024investigating,gonzalez2024effect}.
Finally, two baselines operate in the NAC latent space.
\textit{NAC-SE} is trained to predict clean speech latents given noisy speech latents in a discriminative manner with an $\ell_2$ loss.
It uses a frame-DiT with hidden dimension 768 and no adaLN conditioning.
\textit{EDM-NAC-SE} performs EDM-style diffusion in the low-dimensional codebook lookup space.
It uses a similar frame-DiT conditioned on both the low-dimensional noisy speech latents and the diffusion noise level via adaLN.

\section{Results}

\begin{table*}[t]
\centering
\scriptsize
\setlength{\tabcolsep}{4pt}
\caption{Average objective metrics for each model and each dataset.}
\label{tab:god_table}
\begin{tabular}{l S[table-format=3] S[table-format=1.2] S[table-format=1.2] S[table-format=2.2] S[table-format=1.2] S[table-format=1.2] S[table-format=1.2] S[table-format=1.2] S[table-format=1.2] S[table-format=1.2] S[table-format=1.2] S[table-format=1.2] S[table-format=1.2] S[table-format=2.2] S[table-format=1.2] S[table-format=1.2] S[table-format=1.2] S[table-format=1.2] S[table-format=1.2] S[table-format=1.2] S[table-format=1.2]}
\toprule
& & \multicolumn{10}{c}{Libri-TUT} & \multicolumn{10}{c}{Clarity-FSD50K} \\
\cmidrule(lr){3-12} \cmidrule(lr){13-22}
 & {\rotatebox{90}{Params.\ (M)}} & {\rotatebox{90}{PESQ $\uparrow$}} & {\rotatebox{90}{ESTOI $\uparrow$}} & {\rotatebox{90}{SDR $\uparrow$}} & {\rotatebox{90}{MCD $\downarrow$}} & {\rotatebox{90}{DNSMOS $\uparrow$}} & {\rotatebox{90}{NISQA $\uparrow$}} & {\rotatebox{90}{UTMOS $\uparrow$}} & {\rotatebox{90}{SCOREQ $\uparrow$}} & {\rotatebox{90}{LPS $\uparrow$}} & {\rotatebox{90}{SBS $\uparrow$}} & {\rotatebox{90}{PESQ $\uparrow$}} & {\rotatebox{90}{ESTOI $\uparrow$}} & {\rotatebox{90}{SDR $\uparrow$}} & {\rotatebox{90}{MCD $\downarrow$}} & {\rotatebox{90}{DNSMOS $\uparrow$}} & {\rotatebox{90}{NISQA $\uparrow$}} & {\rotatebox{90}{UTMOS $\uparrow$}} & {\rotatebox{90}{SCOREQ $\uparrow$}} & {\rotatebox{90}{LPS $\uparrow$}} & {\rotatebox{90}{SBS $\uparrow$}} \\
\midrule
Noisy & {--} & 1.31 & 0.61 & 4.96 & 6.76 & 2.92 & 2.03 & 2.44 & 2.47 & 0.82 & 0.73 & 1.34 & 0.74 & 4.96 & 7.52 & 2.90 & 2.46 & 2.42 & 2.46 & 0.82 & 0.69 \\
Clean & {--} & 4.64 & 1.00 & {$\infty$} & 0.00 & 3.73 & 3.31 & 3.06 & 3.98 & 1.00 & 1.00 & 4.64 & 1.00 & {$\infty$} & 0.00 & 3.79 & 3.69 & 3.38 & 4.53 & 1.00 & 1.00 \\
Clean-NAC & 45 & 2.93 & 0.89 & 3.51 & 2.88 & 3.75 & 3.33 & 3.01 & 3.90 & 0.97 & 0.96 & 2.85 & 0.91 & 5.50 & 3.08 & 3.80 & 3.66 & 3.31 & 4.47 & 0.97 & 0.96 \\
\midrule
Conv-TasNet & 5 & {\underline{2.44}} & {\underline{0.85}} & {\underline{15.88}} & 5.03 & 3.58 & 3.27 & 2.68 & 3.49 & {\underline{0.93}} & 0.85 & 2.54 & 0.90 & {\underline{18.18}} & 5.00 & 3.47 & 3.32 & 2.92 & 3.69 & 0.91 & 0.84 \\
BSRNN & 13 & {\bfseries 2.77} & {\bfseries 0.88} & {\bfseries 16.77} & 5.23 & {\bfseries 3.79} & {\bfseries 3.59} & {\bfseries 3.16} & {\bfseries 3.94} & {\bfseries 0.96} & {\bfseries 0.89} & {\bfseries 3.11} & {\bfseries 0.94} & {\bfseries 20.97} & {\underline{4.39}} & 3.72 & {\bfseries 3.74} & {\bfseries 3.32} & {\bfseries 4.22} & {\bfseries 0.95} & {\bfseries 0.88} \\
NAC-SE & 87 & 2.04 & 0.79 & 2.61 & {\underline{4.60}} & 3.69 & 3.15 & 2.97 & {\underline{3.80}} & 0.87 & 0.84 & 2.07 & 0.83 & 4.64 & 4.49 & 3.69 & 3.27 & 3.23 & 4.17 & 0.85 & 0.84 \\
SGMSE+ & 66 & 1.96 & 0.77 & 11.57 & 5.08 & 3.36 & 2.66 & 2.52 & 3.00 & 0.91 & 0.83 & 2.31 & 0.89 & 13.71 & 4.87 & 3.47 & 3.17 & 2.93 & 3.48 & 0.90 & 0.81 \\
EDM-SE & 63 & 2.38 & 0.83 & 13.69 & {\bfseries 4.34} & 3.75 & 3.37 & {\underline{3.02}} & 3.67 & 0.92 & {\underline{0.87}} & {\underline{2.80}} & {\underline{0.92}} & 17.68 & {\bfseries 4.02} & 3.75 & {\underline{3.68}} & {\underline{3.29}} & {\underline{4.20}} & {\underline{0.92}} & {\underline{0.88}} \\
EDM-NAC-SE & 130 & 1.54 & 0.69 & 0.56 & 6.06 & 3.70 & 3.38 & 2.69 & 3.10 & 0.73 & 0.74 & 1.55 & 0.74 & 2.30 & 6.02 & 3.72 & 3.61 & 3.00 & 3.74 & 0.71 & 0.76 \\
\midrule
ADDSE-XS & 4 & 1.48 & 0.66 & 0.55 & 6.31 & 3.67 & 3.38 & 2.66 & 2.86 & 0.65 & 0.70 & 1.46 & 0.71 & 2.10 & 6.27 & 3.67 & 3.58 & 2.90 & 3.41 & 0.59 & 0.70 \\
ADDSE-S & 17 & 1.56 & 0.69 & 0.76 & 6.04 & 3.72 & 3.40 & 2.78 & 3.18 & 0.71 & 0.73 & 1.56 & 0.73 & 2.44 & 6.03 & 3.72 & 3.61 & 3.06 & 3.77 & 0.67 & 0.74 \\
ADDSE-M & 65 & 1.62 & 0.70 & 0.89 & 5.95 & 3.74 & 3.41 & 2.88 & 3.38 & 0.76 & 0.76 & 1.62 & 0.75 & 2.65 & 5.87 & 3.74 & 3.62 & 3.14 & 3.96 & 0.73 & 0.76 \\
ADDSE-L & 259 & 1.66 & 0.71 & 1.01 & 5.85 & 3.75 & 3.41 & 2.89 & 3.46 & 0.78 & 0.77 & 1.67 & 0.76 & 2.79 & 5.75 & {\underline{3.76}} & 3.63 & 3.17 & 4.05 & 0.75 & 0.78 \\
ADDSE-XL & 580 & 1.68 & 0.72 & 1.10 & 5.83 & {\underline{3.76}} & {\underline{3.43}} & 2.90 & 3.52 & 0.79 & 0.78 & 1.70 & 0.77 & 2.88 & 5.69 & {\bfseries 3.76} & 3.64 & 3.20 & 4.10 & 0.77 & 0.79 \\
\bottomrule
\end{tabular}
\raggedright
Values in \textbf{bold} indicate best performance excluding Noisy, Clean, and Clean-NAC.
Second best is \underline{underlined}.
\end{table*}

\begin{figure}[ht!]
  \centering
  \includegraphics[width=0.994\linewidth]{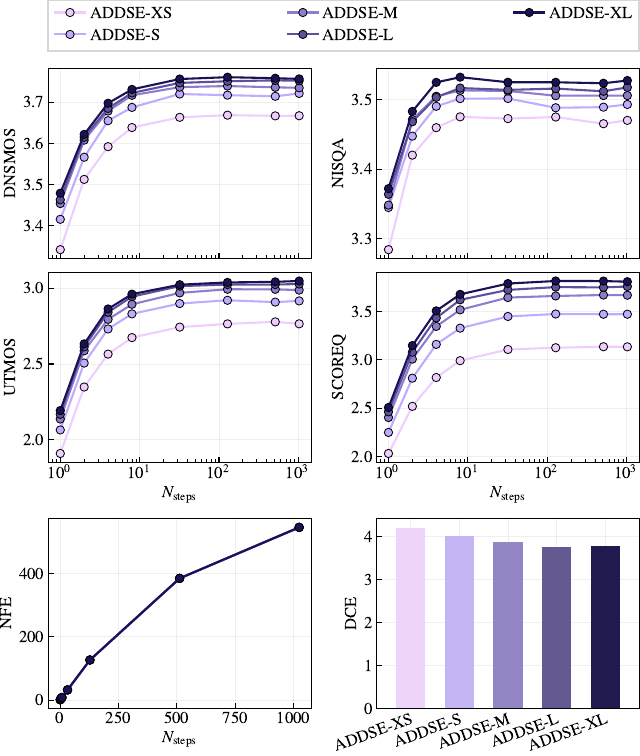}
  \caption{
    Top and middle rows: Non-intrusive metrics as a function of $N_\text{steps}$.
    Bottom left: Number of function evaluations as a function of $N_\text{steps}$.
    Bottom right: DCE on test data.
  }
  \label{fig:addse_nsteps_both}
\end{figure}

\begin{figure}[ht!]
  \centering
  \includegraphics[width=\linewidth]{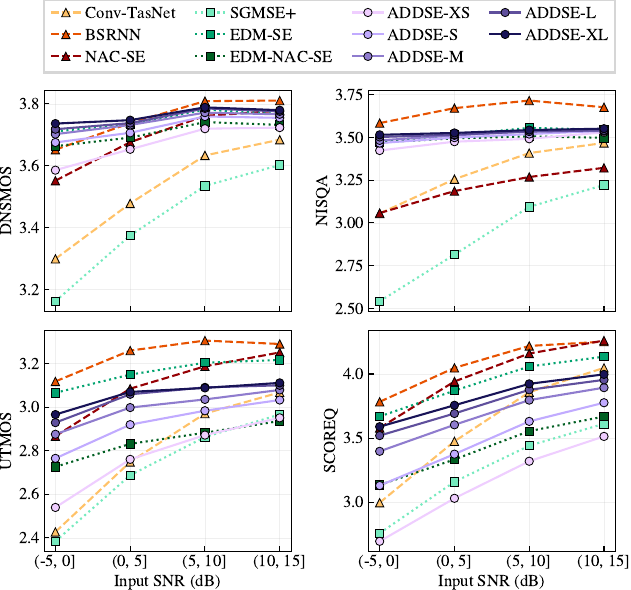}
  \caption{Non-intrusive metrics as a function of input SNR.}
  \label{fig:sdr_bins_both}
\end{figure}

Systems are compared using 4 intrusive metrics PESQ~\cite{rix2001perceptual}, ESTOI~\cite{jensen2016algorithm}, SDR, and MCD~\cite{kubichek1993mel}, 4 non-intrusive metrics DNSMOS~\cite{reddy2022dnsmos}, NISQA~\cite{mittag2021nisqa}, UTMOS~\cite{baba2024t05}, and SCOREQ~\cite{ragano2024scoreq}, and 2 downstream-task-independent metrics LPS~\cite{pirklbauer2023evaluation} and SBS~\cite{saeki2024speechbertscore}.
Average results for each model and each dataset are shown in Table~\ref{tab:god_table}.
\textit{Clean-NAC} denotes the clean speech encoded and decoded by the NAC.
For the ADDSE systems, the best $N_\text{steps}$ is selected.
The NAC-based systems achieve poor PESQ, ESTOI and SDR, which is expected since the NAC does not reconstruct the clean speech phase.
However, they achieve competitive non-intrusive metrics.
For example, all ADDSE systems, including the small 4\,M-parameter ADDSE-XS, outperform Conv-TasNet and SGMSE+ on DNSMOS and NISQA for both datasets.
Moreover, ADDSE-XL shows the best DNSMOS on Clarity-FSD50K, and the second best NISQA on both datasets.
However, BSRNN and EDM-SE show the best performance in most cases.

Figure~\ref{fig:addse_nsteps_both} shows the non-intrusive metrics averaged across both datasets as a function of $N_\text{steps}$ for the ADDSE systems.
The best NISQA is achieved at $N_\text{steps} = 8$, while the remaining metrics plateau at $N_\text{steps} = 16$, suggesting good performance is achievable with few sampling steps.
Also shown is the effective number of function evaluations (NFE) as a function of $N_\text{steps}$.
For $N_\text{steps} \geq 128$, NFE is lower than $N_\text{steps}$ for all systems.
This is because the probabilities predicted by the network can be reused (see Sec.~\ref{sec:add}).
For $N_\text{steps} = 1024$, the average NFE is 545, which represents nearly a 2$\times$ speedup compared to time-dependent modeling.
Finally, the figure shows the DCE~\eqref{eq:dce} averaged over both test datasets.
The DCE decreases as model size increases, which correlates with the non-intrusive metrics.

Figure~\ref{fig:sdr_bins_both} shows the non-intrusive metrics averaged across both datasets as a function of the input SNR for all systems.
The benefit of ADDSE over Conv-TasNet and SGMSE+ is more pronounced at low input SNRs.
Notably, ADDSE shows high robustness to input SNRs on DNSMOS and NISQA.
We attribute this to the strong speech prior learned by the NAC, which allows producing high-quality speech even in very noisy conditions.

\section{Conclusion}

We proposed ADDSE, a framework for SE based on ADD in the latent space of a NAC.
The discrete nature of the speech codes allows reusing the network predictions for multiple sampling steps, leading to more efficient sampling compared to continuous diffusion.
We also proposed RQDiT, a non-autoregressive architecture for modeling the hierarchical structure of RVQ codes.
Results on two datasets show competitive performance as measured by non-intrusive metrics, especially at low input SNRs and with few sampling steps.
Future work includes extending to full-band SE and incorporating semantic coding.

\bibliographystyle{IEEEtran}
\bibliography{mybib}

\end{document}